\documentclass{article}

\usepackage{microtype}
\usepackage{graphicx}
\usepackage{subfigure}
\usepackage{booktabs} 
\usepackage{multirow}
\usepackage{tabularx}
\usepackage{graphicx}
\usepackage{subcaption}
\usepackage{enumitem}

\usepackage[ruled,vlined]{algorithm2e}
\SetKwComment{Comment}{$\triangleright$\ }{}
\SetKwInput{KwInput}{Input}  
\SetKwInput{KwOutput}{Output}
\SetKwInput{KwInit}{Initialize}
\SetAlgoBlockMarkers{begin}{end}

\usepackage{hyperref}

\usepackage[accepted]{icml2025}

\usepackage{amsmath}
\usepackage{amssymb}
\usepackage{mathtools}
\usepackage{amsthm}
\usepackage{tcolorbox}

\usepackage[capitalize,noabbrev]{cleveref}

\theoremstyle{plain}

\theoremstyle{definition}

\theoremstyle{remark}

\usepackage[textsize=tiny]{todonotes}
\usepackage{changepage}

\def\BibTeX{{\rm B\kern-.05em{\sc i\kern-.025em b}\kern-.08em
    T\kern-.1667em\lower.7ex\hbox{E}\kern-.125emX}}

\begin{document}

\twocolumn[
\icmltitle{Hierarchical Debate-Based Large Language Model (LLM) for Complex Task Planning of 6G Network Management}

\icmlsetsymbol{equal}{*}

\begin{icmlauthorlist}
\icmlauthor{Yuyan Lin}{equal,yyy}
\icmlauthor{Hao Zhou}{equal,yyy}
\icmlauthor{Chengming Hu}{yyy}
\icmlauthor{Xue Liu}{yyy}
\icmlauthor{Hao Chen}{comp}
\icmlauthor{Yan Xin}{comp}
\icmlauthor{Jianzhong (Charlie) Zhang}{comp}
\end{icmlauthorlist}

\icmlaffiliation{yyy}{School of Computer Science, McGill University, Montreal, QC H3A 0E9, Canada.}
\icmlaffiliation{comp}{Samsung Research America, Plano, Texas, TX 75023, USA.}
\icmlcorrespondingauthor{Hao Zhou}{haozhou029@gmail.com. The 6GPlan dataset is available at \url{https://github.com/haozhou1995/6GPlan_Dataset.git}}

\vskip 0.3in
]
\printAffiliationsAndNotice{\icmlEqualContribution}

\thispagestyle{fancy}            
\chead{\textbf{Accepted by 2025 International Conference on Machine Learning (ICML) Workshop on ML4Wireless}} 

\renewcommand{\headrulewidth}{0pt}      
\pagestyle{plain}

\begin{abstract}
6G networks have become increasingly complicated due to novel network architecture and newly emerging signal processing and transmission techniques, leading to significant burdens to 6G network management. 
Large language models (LLMs) have recently been considered a promising technique to equip 6G networks with AI-native intelligence. 
Different from most existing studies that only consider a single LLM, this work involves a multi-LLM debate-based scheme for 6G network management, where multiple LLMs can collaboratively improve the initial solution sequentially. 
Considering the complex nature of 6G domain, we propose a novel hierarchical debate scheme: LLMs will first debate the sub-task decomposition, and then debate each sub-task step-by-step. 
Such a hierarchical approach can significantly reduce the overall debate difficulty by sub-task decomposition, aligning well with the complex nature of 6G networks and ensuring the final solution qualities.
In addition, to better evaluate the proposed technique, we have defined a novel dataset named 6GPlan, including 110 complex 6G network management tasks and 5000 keyword solutions.   
Finally, the experiments show that the proposed hierarchical debate can significantly improve performance compared to baseline techniques, e.g. more than 30\% coverage rate and global recall rate improvement.

\end{abstract}

\begin{figure*}[t]
\centering
\includegraphics[width=0.95\linewidth]{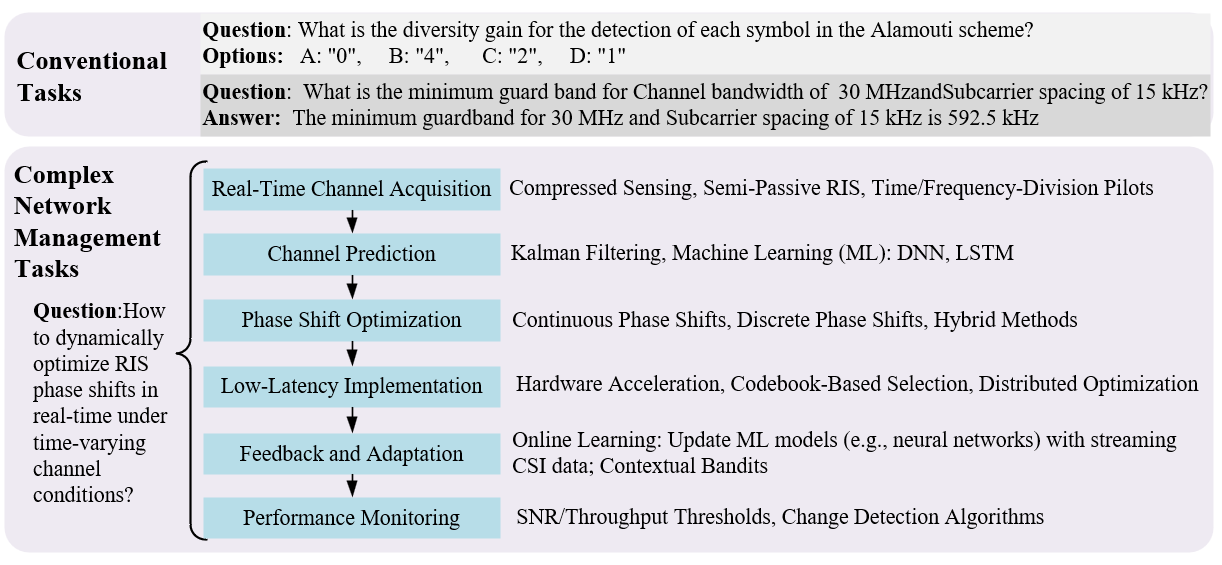}
\caption{Comparisons between conventional multiple-choice problems and our considered complex network management tasks.}
\label{fig-mana}
\end{figure*}

\section{Introduction}

The envisioned 6G networks are expected to incorporate many advanced paradigms, such as terahertz (THz) communications, reconfigurable intelligent surfaces (RIS), cell-free massive MIMO, semantic communication, and so on, and the resulting heterogeneity will greatly increase overall system complexity~\cite{giordani2020toward}.
Meanwhile, the interaction between these novel techniques also poses unprecedented challenges for 6G network management, e.g., RIS phase configurations influence beamforming decisions, and machine learning-enabled scheduling algorithms must adapt to rapidly varying THz channel conditions~\cite{zhou2023survey}. 
Therefore, given such complexity, ensuring end-to-end performance, reliability, and energy efficiency in 6G networks can be challenging.
Specifically, 6G network management demands a thorough understanding of network architecture, professional knowledge of various 6G techniques, and the capability of integrating cross-layer optimization frameworks.
Academic studies have proposed diverse solutions to optimize 6G networks, but most of them are limited to a single algorithm and architecture~\cite{shi2023machine}.  
By contrast, practical network management usually requires cross-layer optimization, monitoring, reporting, and continuous improvement. It indicates a significant gap between proof-of-concept algorithm solutions and real-world network management implementations.

Given the above challenges and difficulties, large language models (LLMs) have recently been considered promising solutions for complicated network management tasks~\cite{zhou2024large}. 
LLMs can equip 6G network management with rich real‐world knowledge such as vendor manuals, standards specifications, and historical operational logs. Consequently, they can not only understand the intricate details of each network element, but also retain a holistic view of system behaviour. 
Such foundations also enable LLMs to offer clear, human-centred explanations for their recommendations and decisions, fostering network operator trust and transparent decision making~\cite{zhou2024large2}. 
Moreover, by reasoning across diverse network data domains, LLMs can orchestrate cross‐layer optimizations, e.g, generating comprehensive solutions for network management and coordinating multiple layers.

Existing studies have explored various applications of LLM-enabled 6G networks, including network optimization~\cite{zhou2024large2}, traffic prediction~\cite{hu2024self}, network security~\cite{nguyen2024large}, federated LLM ~\cite{su2024federated,yan2025federated}, semantic communication~\cite{park2024towards,chen2024personalizing}, etc. 
These studies have demonstrated the great potential of LLMs to improve network intelligence and efficiency in the 6G era.  
Note that most of these works rely on the knowledge of a single LLM to understand the defined network problem and generate outputs. 
However, a single LLM’s knowledge is limited to the corpora it was trained on, which may under-represent the latest standards updates, region-specific operational practices, leading to outdated generated content~\cite{lu2024merge}.
Additionally, without domain-specific grounding, a standalone LLM can invent parameter names or protocol behaviours, undermining the reliability of decision-making and preventing the applications to crucial 6G scenarios.

\begin{figure*}[t]
\centering
\includegraphics[width=1\linewidth]{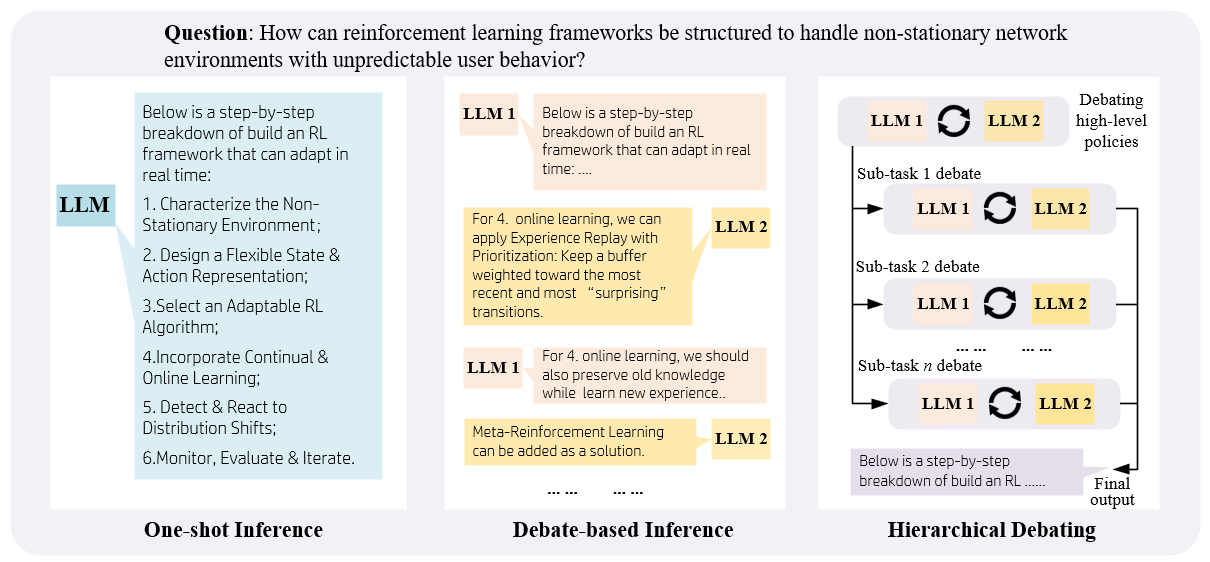}
\caption{Comparisons between conventional one-shot inference, debating and hierarchical debate.}
\label{fig-all}
\vspace{-15pt}
\end{figure*}

To this end, this work considers the collaboration of multiple LLMs to address complex network management tasks in 6G domain. 
Introducing multi-LLM collaboration can overcome the inherent limitations of a single LLM, improving the reliability and reducing hallucinations.  
In particular, given a specific network management task, we encourage multiple LLMs to improve the initial solutions iteratively, e.g., completing the former solutions or proposing alternatives. 
Such a debate-based approach has been studied in multiple existing studies, including diverge thinking~\cite{liang2023encouraging}, theoretical analyses~\cite{estornell2024multi}, round-robin
style debate~\cite{chan2023chateval}, and using judges for evaluation~\cite{khan2024debating}, etc.  
These studies demonstrate that LLM debating can significantly improve the generated content quality.

However, network management in 6G is far more complex than a simple multiple‐choice task, since it demands end-to-end orchestration across data collection, predictive analytics, optimization loops, continuous improvement, etc~\cite{tshakwanda2024advancing}.
For instance, the optimization loop is central to many optimization tasks, and the corresponding algorithms include convex optimization, reinforcement learning, meta-heuristic algorithms, etc~\cite{zhou2023survey}. 
In addition, a closed-loop feedback mechanism is usually required to validate the impact of each action, retrain the predictive models, and update optimization policies, making the overall pipeline a tightly coupled, multi-stage workflow rather than a one-off inference.

Therefore, considering the unique demand of complex network management tasks, the unique contribution of this work is summarized by:
1) Firstly, we propose a novel hierarchical debate framework. 
In particular, LLMs will first focus on task decomposition, decoupling complex network management tasks into more actionable sub-tasks. 
Then, LLMs can concentrate on each sub-task, proposing/improving specific pipelines and relevant techniques. 
Compared with regular debating in existing studies~\cite{liang2023encouraging, estornell2024multi, chan2023chateval, khan2024debating}, such a hierarchical design can reduce the overall debate complexity, since each debate can focus on more specific sub-tasks, instead of improving the whole problem in a single operation.
In addition, decoupling the whole problem into multiple sub-tasks allows for parallel execution, saving the overall implementation time. 
\\
2) To better evaluate the performance of the proposed hierarchical debate techniques, we further build a complex network management task planning dataset named 6GPlan, including eleven 6G-related techniques such as reconfigurable intelligent surfaces, Open RANs, quantum communication, semantic communication, etc. It involves 110 complicated planning/management tasks in the 6G domain, and the solutions consist of around 5,000 keywords. 
Compared with existing studies that focus on one specific topic~\cite{zhou2024large2, hu2024self, nguyen2024large, su2024federated}, 
this work is the first to utilize LLMs to comprehensively understand complex 6G network management problems.  
Finally, the experiments show that the proposed hierarchical debate can significantly improve performance compared to baseline techniques, e.g. more than 30\% coverage rate and global recall rate improvement.

\section{LLM-Based Complicated Network Task Planning and Management}

Fig.\ref{fig-mana} shows the differences between conventional problems and our considered complex network management problems. 
Specifically, these conventional problems only require one-off inference. For instance, multiple-choice problems are defined in most existing datasets to evaluate the performance of LLMs on telecom knowledge understanding, e.g., TeleQnA\cite{maatouk2023teleqna}, NetEval\cite{miao2023empirical}, and ORAN-bench-13 \cite{gajjar2025oran} . Consequently, the LLM can simply recall a known fact, then match it to one of the given options. 

By contrast, our considered complex network management tasks are open-ended and require LLMs to develop a solution from scratch. 
Firstly, from the reasoning and planning perspective, the ideal solution is expected to integrate several processing stages: channel acquisition and prediction, phase-shift optimization, implementation, feedback, and performance monitoring, demonstrating true chain-of-thought capability. For instance, due to the “real-time” requirements in the question, the solutions in Fig. \ref{fig-mana} highlight “\textit{Real-time Channel Acquisition}” and “\textit{Low-latency Implementation}”. 
Secondly, in terms of the depth and breadth of knowledge, such open-ended tasks force LLMs to draw on a broad base of telecom domain facts, including signal processing fundamentals, wireless channel modelling, quantization effects, hardware constraints, and finally combine them into a coherent pipeline. For example, LLMs have to consider different RIS designs in Fig. \ref{fig-mana}, e.g., “\textit{Continuous Phase Shifts}”, “\textit{Discrete Phase Shifts}”, and “\textit{Hybrid Methods}”.
Moreover, testing LLMs on multi-disciplinary workflows also allows us to assess the capacity to integrate heterogeneous methods rather than retrieve isolated facts, e.g., combining “\textit{Compressed Sensing}”, “\textit{DNNs/LSTMs}”, “\textit{Hardware Acceleration}”, and “\textit{Online Learning}” techniques.

The above analyses demonstrate that handling complex network management tasks demands dynamic, decompositional problem-solving, integration of heterogeneous techniques, and real-time adaptation.
Therefore, compared with conventional multiple‐choice or Q\&A tasks, complex network management scenarios provide a far more demanding and informative benchmark for LLMs.

\section{Hierarchical debate-Based Task Planning}

This section will first introduce conventional debate-based network management task planning, and then present the proposed hierarchical debate-based framework along with the designed 6GPlan dataset. 

\subsection{Debate-based Task Planning}
\label{sec-debate}
Fig. \ref{fig-all} compares one-shot inference, debate-based inference, and the proposed hierarchical debate scheme. 
Specifically, given a complex open-ended question, the conventional approach will generate the final solution directly by one-shot inference. 
By contrast, the debate-based inference involves multiple LLMs, and these LLMs can improve/criticize the former solutions iteratively.

\textbf{Meta Prompts:} In the meta prompt, we briefly introduce the task and rules: 
\begin{adjustwidth}{0.3cm}{0.3cm}
\textit{“We are in an iterative debate process. Given a technical question \{\texttt{Question}\} and category \{\texttt{6G\_Category}\}, we aim to generate better solutions by improving the initial results iteratively.”}
\end{adjustwidth}

\textbf{Debater Prompts:} Then, for these complex network management tasks, instead of criticizing the former solutions, we encourage the LLM to improve the former solutions by adding new related techniques or alternatives to existing techniques. 
\begin{adjustwidth}{0.3cm}{0.3cm}
\textit{“You are a expert in \{\texttt{6G\_Category}\} research. Based on the given question \{\texttt{Question}\} and previous solutions \{\texttt{Pre\_Solution}\}, please improve it by adding any missing technical keywords, methods, or alternative approaches.”}
\end{adjustwidth}

For instance, as shown in Fig. \ref{fig-all}, LLM 2 can complete the initial solutions proposed by LLM 1 by adding \textit{“Experience Replay with Prioritization”}, and then LLM 1 can further improve it by claiming \textit{“we should also preserve old knowledge while learn new experience”}. 
Finally, note that no judge is needed for the defined scheme, since it aims to improve the solution iteratively instead of judging correct/wrong answers.

\begin{algorithm}[t]
\caption{Hierarchical Debate (Sequential Debaters)}
\label{alg:hierarchical_debate_seq}

\KwInput{Technical question $Q$, 6G research category $C$, Number of debaters $M$} 
\KwOutput{Final refined solutions for each sub-task}
\KwInit{High-level steps $S \gets S^{(0)}$ (empty or seed steps)} 
\textbf{Phase 1: Task Decomposition Debate} \\
\For{$i \gets 1$ \KwTo $N_{\mathrm{decomp}}$}{
  $S_{\mathrm{cur}} \gets S$\\ 
  \For{$j \gets 1$ \KwTo $M$}{
    \textbf{Meta-Prompt:} Iterative debate. Given $Q$ and $C$, decompose into sub-tasks by improving $S_{\mathrm{cur}}$ \\ 
    \textbf{Debater $j$ Prompt:} Expert in $C$. Based on $Q$ and $S_{\mathrm{cur}}$, improve the sub-tasks \\ 
    $S_{\mathrm{cur}} \gets$ output of Debater $j$
   }
   \textbf{Endfor}\\
  $S \gets S_{\mathrm{cur}}$
}
\textbf{Endfor}\\
Let $S^* \gets S$ (final decomposition)

\vspace{1ex}
\textbf{Phase 2: Sub-Task Implementation Debate} \\
\For{each sub-task $s_n$ in $S^*$}{
  $R \gets R^{(0)}_n$ \\
  \For{$k \gets 1$ \KwTo $N_{\mathrm{sub}}$}{
    $R_{\mathrm{cur}} \gets R$ \\
    \For{$j \gets 1$ \KwTo $M$}{
      \textbf{Meta-Prompt:} Iterative debate. Given $Q$ and sub-task $s_n$, improve $R_{\mathrm{cur}}$ \\ 
      \textbf{Debater $j$ Prompt:} Expert in $C$. Based on $Q$, $s_n$, and $R_{\mathrm{cur}}$, add missing methods/keywords \\ 
      $R_{\mathrm{cur}} \gets$ output of Debater $j$}
      \textbf{Endfor}\\
    $R \gets R_{\mathrm{cur}}$}
    \textbf{Endfor}\\
  Store final $R^*_n \gets R$}
  \textbf{Endfor}\\
\Return{$\{R^*_n\}_{n=1}^{|S^*|}$}
\label{algo-1}
\end{algorithm}

\subsection{Hierarchical debate}

Considering the inherent complexity of network management tasks, here we propose a novel hierarchical debate method. 
In particular, hierarchical debate includes 2 phases: task decomposition debate and sub-task debate.

1) Firstly, for the \textbf{task decomposition phase}, we aim to decompose the overall task into more specific sub-tasks. 

\begin{adjustwidth}{0.5cm}{0cm}

\textbf{Meta Prompts:} In the meta prompt, we briefly introduce the task and rules to decompose the question into more specific sub-tasks: 
\begin{adjustwidth}{0.3cm}{0.3cm}
\textit{“We are in an iterative debate process. Given a technical question \{\texttt{Question}\} and category \{\texttt{6G\_Category}\}, we aim to decompose the question into sub-tasks by iteratively improving the initial technical steps.”}
\end{adjustwidth}

\textbf{Debater Prompts:} The debate prompt aims to review and improve the high-level technical steps. 
\begin{adjustwidth}{0.3cm}{0.3cm}
\textit{“You are a expert in \{\texttt{6G\_Category}\} research. 
Based on the given question \{\texttt{Question}\} and previous task decomposition results \{\texttt{Technical\_steps}\}, please serve as a critical reviewer to improve the sub-task steps.”}
\end{adjustwidth}
\end{adjustwidth}

2) Then, for the sub-task implementation phase, we enforce LLMs to debate on each sub-task.

\begin{adjustwidth}{0.5cm}{0cm}
\textbf{Meta Prompts:} In the meta prompt, we briefly introduce the task and rules, and ask the LLM to focus exclusively on a specific sub-task: 
\begin{adjustwidth}{0.3cm}{0.3cm}
\textit{“We are in an iterative debate process. Given a technical question \{\texttt{Question}\} and category \{\texttt{Category}\}, please focus exclusively on sub-task \{\texttt{Step\_n}\} regarding the question. We aim to improve the initial results iteratively”.}
\end{adjustwidth}

\textbf{Debater Prompts}: Similar to the regular debate process in Section \ref{sec-debate}, LLMs are encouraged to improve the previous solutions.

\begin{adjustwidth}{0.3cm}{0.3cm}
\textit{“You are a expert in \{\texttt{6G\_Category}\} research. Based on the given question \{\texttt{Question}\}, please focus on sub-task \{\texttt{Step\_n}\} exclusively and improve the previous solutions \{\texttt{Pre\_Solution}\} of this sub-task by adding any missing technical keywords, methods, or alternative approaches.”}
\end{adjustwidth}

\end{adjustwidth}

Finally, the proposed hierarchical debate is summarized in Algorithm \ref{algo-1}. This two-level, sequential-debater design ensures that a) high-level planning is solidified before b) detailed solutions are honed. A detailed high-level plan is expected to guide the low-level debate, producing more comprehensive solutions.

\subsection{Dataset Design}

\begin{figure}[t]
\centering
\includegraphics[width=1\linewidth]{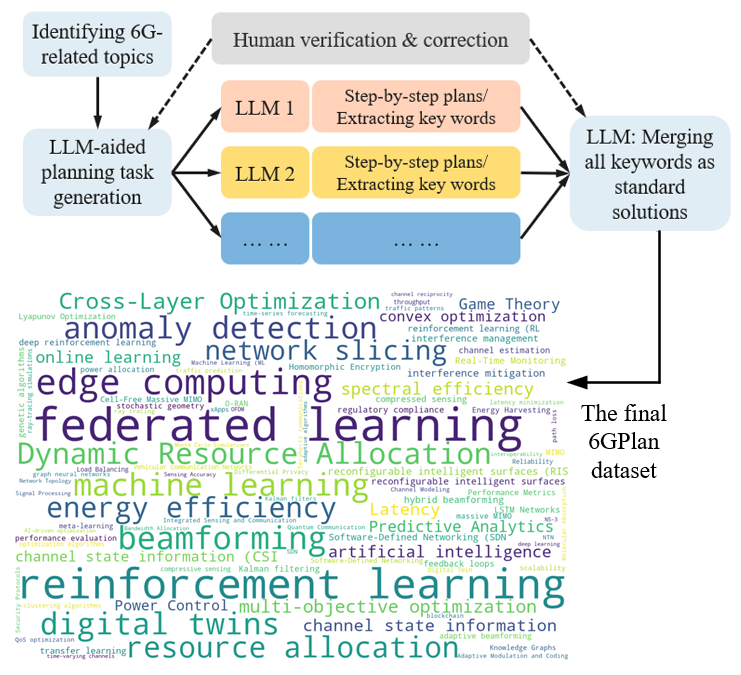}
\caption{The design pipeline of 6GPlan dataset.}
\label{fig-data}
\vspace{-15pt}
\end{figure}

Fig. \ref{fig-data} shows the overall design pipeline of the 6GPlan dataset\footnote{The 6GPlan dataset is available at \url{https://github.com/haozhou1995/6GPlan_Dataset.git}}. 
Firstly, we selected 11 topics regarding 6G networks, e.g., integrated sensing and communication, mmWave and Terahertz Communications, 
non-terrestrial networks, cell-free massive MIMO, etc. 
For each category, we asked an LLM to generate related questions, focusing on complex network management and optimization tasks. 
After that, we consider a multi-LLM question-answering approach: asking multiple LLMs to generate solutions for a given question, and then extracting related technical keywords from their replies. 
Here we use these keywords to represent the key elements that should be covered in a high-quality solution. 
At the end, we will merge the keywords from these LLMs, and implement human verification and correction, guaranteeing the dataset quality.

\begin{table*}[t]
  \centering
  \caption{Experimental results across pipelines and model combinations.}
  \label{tab:exp_results}
  \small
  \setlength{\tabcolsep}{6pt}
  \begin{tabularx}{\textwidth}{lXcccc}
    \toprule
    \textbf{Pipeline} 
      & \textbf{Model Combination} 
      & \textbf{MCR (\%)} 
      & \textbf{KHC} 
      & \textbf{GRR (\%)}  \\
    \midrule
    \multirow{5}{*}{Baseline}
      & GPT-4o                     & 36.99            & 34.01   & 36.57  \\
      & GPT-4o-mini                & 39.62            & 36.58   & 39.34  \\
      & GPT-3.5-turbo              & 22.79            & 21.00   & 22.58  \\
      & LLaMA3-8B                  & 31.42            & 28.94   & 31.12  \\
    \midrule
    \multirow{3}{*}{Regular Debate}
      & GPT-4o + GPT-3.5-turbo     & 38.65            & 33.05   & 35.55  \\
      & GPT-4o + GPT-4o-mini       & 49.75            & 45.87   & 49.33  \\
      & LLaMA3-8B + GPT-3.5-turbo  & 39.86            & 35.56   & 38.24  \\
    \midrule
    \multirow{3}{*}{Hierarchical Debate}
      & GPT-4o + GPT-3.5-turbo     & 55.22            & 51.05   & 54.90   \\
      & GPT-4o + GPT-4o-mini       & \textbf{81.19}   & \textbf{75.41}   & \textbf{81.09}  \\
      & LLaMA3-8B + GPT-3.5-turbo  & 58.30            & 55.02   & 59.17    \\
    \bottomrule
  \end{tabularx}
  \vspace{-15pt}
\end{table*}

\begin{figure*}[!t]
  \centering
  \subfigure[GPT-4o + GPT-3.5-turbo]{%
    \includegraphics[width=0.33\textwidth,height=0.3\textheight,keepaspectratio]{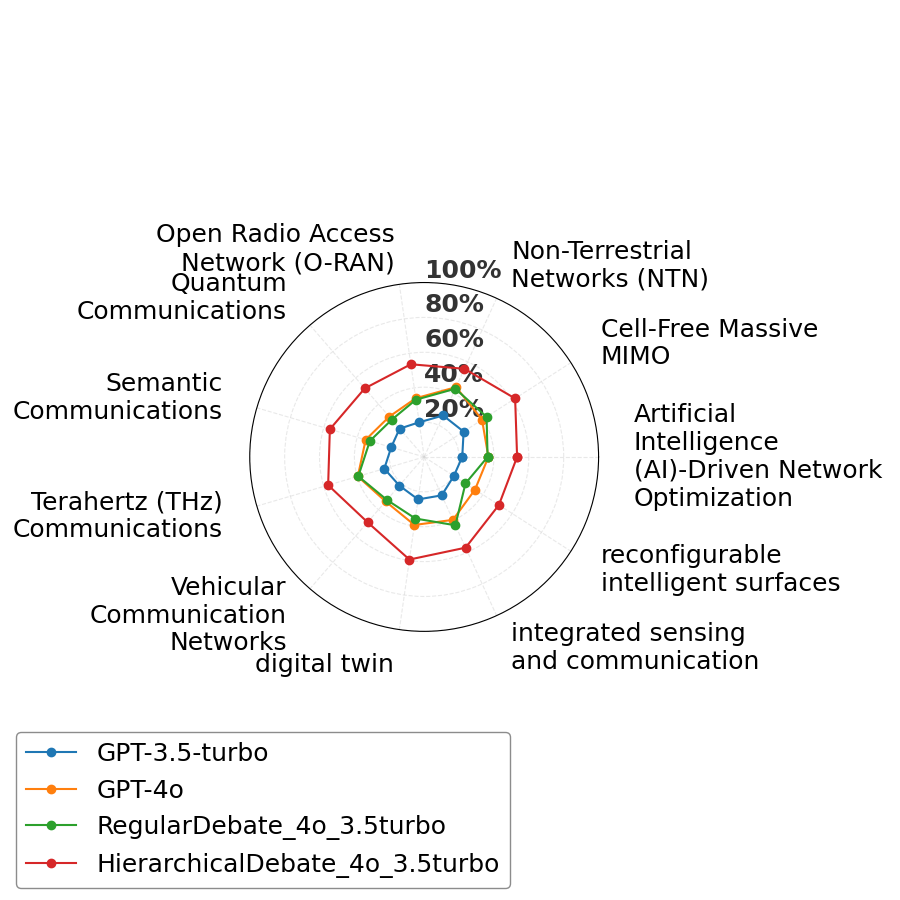}%
    \label{fig:radar-4o-3.5-turbo}%
  }
  \hfill
  \subfigure[GPT-4o + GPT-4o-mini]{%
    \includegraphics[width=0.33\textwidth,height=0.3\textheight,keepaspectratio]{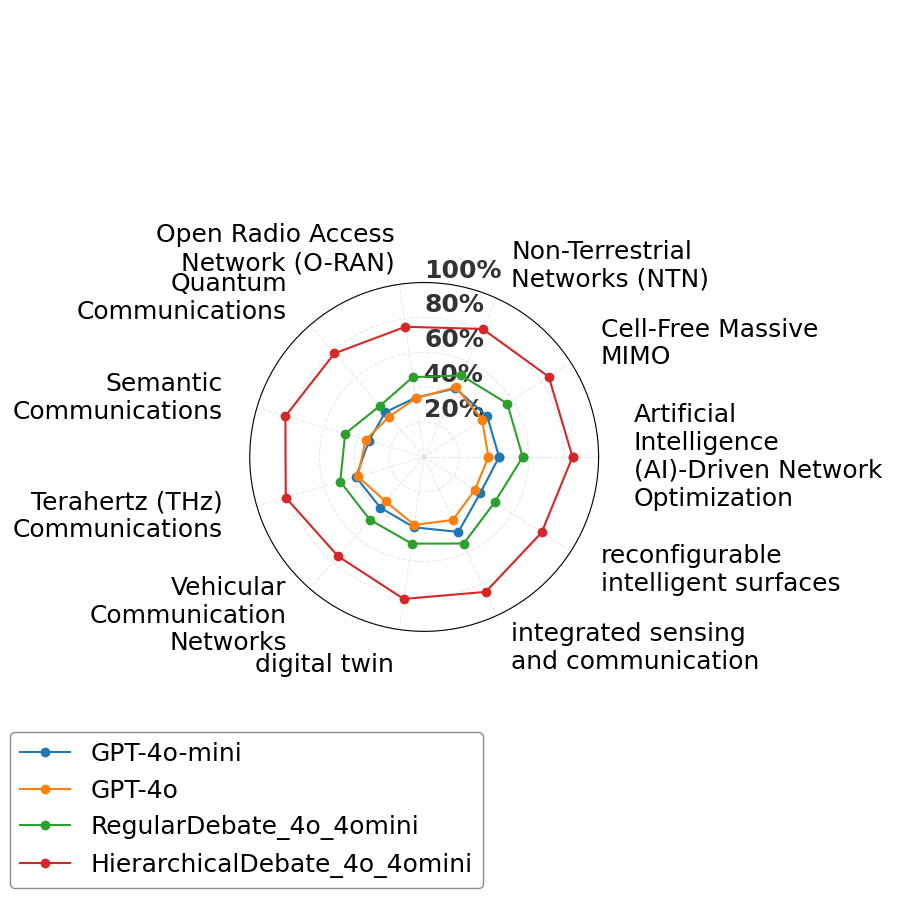}%
    \label{fig:radar-4o-4omini}%
  }
  \hfill
  \subfigure[GPT-3.5-turbo + Llama3-8B]{%
    \includegraphics[width=0.33\textwidth,height=0.3\textheight,keepaspectratio]{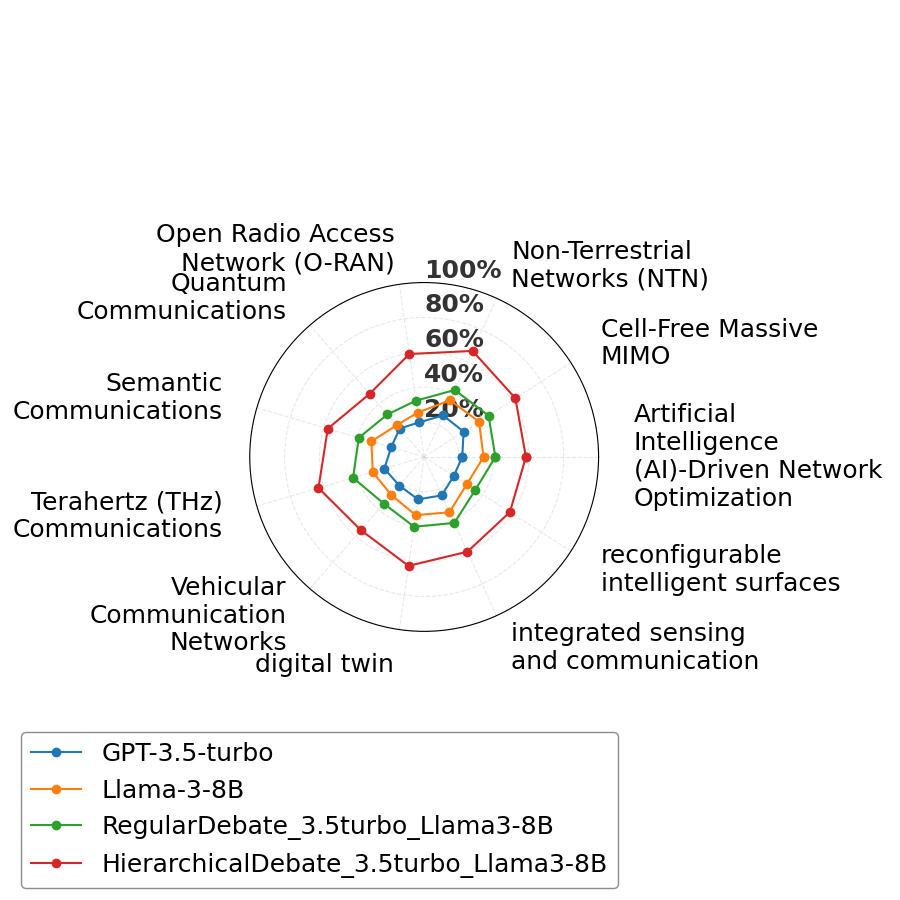}%
    \label{fig:radar-3.5-turbo-3-8b}%
  }

  \caption{Radar‐plot comparison of MCR for three debate configurations over 11 6G technical categories.}
  \label{fig:radar-combined}
    \vspace{-15pt}
\end{figure*}

\section{Performance Evaluation}

\subsection{Experiment Settings}

We evaluated all experiments on the 6GPlan dataset, which contains 110 complicated planning/management tasks distributed across 11 core 6G themes (e.g., RIS, semantic communications, mmWave/Terahertz networks). Each question is paired with a set of gold-standard technical keywords (around 5000 total) that serve as our reference for evaluation. 
Detailed dataset samples can be found in the Appendix \ref{datasetsample-app}.
Experiments ran concurrently on 16 worker threads, with a 1 s delay between API calls to avoid rate-limit failures.

We evaluated three multistage solution generation pipelines: Baseline, Regular Debate, and Hierarchical Debate, involving five large language models (LLMs): GPT-3.5-turbo, GPT-4o, GPT-4o-mini, and Llama3-8b.
\newline \textbf{Baseline:}Each model receives a single end-to-end prompt per question that asks for a comprehensive, narrative solution in the form of technical keywords and methods, with no intermediate extraction or extra interaction. \\
\textbf{Regular Debate:} This pipeline is augmented by structuring knowledge elicitation as a round of debate: the first debater proposes an initial solution, and a second debater enriches it by adding missing domain-specific alternatives, algorithms, and designs. We systematically compared three combinations of LLMs, GPT-4o with GPT-4o-mini, GPT-4o with GPT-3.5-turbo, and GPT-3.5-turbo with Llama 3-8B. \\
\textbf{Hierarchical Debate:}  In this multi‐stage pipeline as shown in Algorithm \ref{algo-1}, we first conduct a one‐round top-level debate in which Debater 1 breaks the original question into a numbered list of technical sub-tasks and Debater 2 enriches or adds any missing steps. Each resulting sub-task then undergoes an independent, one-round sub-task-level debate following the Regular Debate protocol: Debater 1 generates an initial solution for that sub-task, and Debater 2 refines it by appending any omitted alternatives or algorithms. Finally, the outputs from all sub-task debates are concatenated into a single, cohesive solution.
See detailed Prompt input of each pipeline in Appendix \ref{sec:prompt_1}.\\
\textbf{Evaluation}. All outputs were evaluated against the gold-standard keywords using 3 complementary metrics—Macro Coverage Rate (MCR) (\%), Keyword Hit Count (KHC), and Global Recall Rate (GRR)(\%).
MCR assesses equity by averaging per-question recall across all prompts, ensuring that performance on shorter or niche items is valued equally to that on longer, information-dense ones.
GRR measures overall retrieval capacity by computing the corpus-wide ratio of matched to reference keywords, thereby reflecting a model’s effectiveness on prompts with extensive keyword inventories.
KHC is the mean number of correctly retrieved keywords per prompt, reflecting a model’s absolute yield and revealing whether it tends to over‐ or under‐generate terms. 
Together, these measures ensure equity (MCR), breadth (GRR), and yield (KHC).


\subsection{Experiment Results}
On the 6GPlan benchmark of 110 open-ended questions, Regular Debate achieved modest gains, improving recall by up to 10 \% over the one-shot Baseline by applying the adversarial debate framework first described by Irving et al. [\cite{irving2018aisafety}]. In contrast, our Hierarchical Debate pipeline which builds on the Iterated Amplification framework proposed by Christiano et al. [\cite{christiano2018supervising}] delivers substantially greater improvements.
By decomposing each question into focused sub-tasks and refining them individually, Hierarchical Debate pushes MCR into the 55–81 \% range—more than doubling Baseline performance in our strongest configuration and outperforming Regular Debate by over 30 \% (See Table \ref{tab:exp_results}). 
Parallel increases in KHC and GRR confirm that this multi‐stage approach not only recovers a larger fraction of reference terms but also significantly raises the absolute volume of keywords retrieved. Therefore, staged decomposition and targeted sub-task debate are essential for achieving comprehensive, reliable coverage in complex technical question answering and complex planning tasks in the 6G domain.

Moreover, Figure \ref{fig:radar-combined} breaks down per‐category MCR for three representative model stacks, GPT-4o+GPT-3.5-turbo, GPT-4o+GPT-4o-mini, and Llama3-8B+GPT-3.5-turbo, and once again confirms the decisive advantage of Hierarchical Debate. 
In every subplot, the Hierarchical Debate trace (red) completely encloses both the one‐shot Baseline (blue) and single‐round Regular Debate (green) polygons, boosting MCR by roughly 20–35 \% across all eleven 6G themes. In particular, underserved domains such as Semantic Communications and Quantum Communications jump from near‐baseline levels of ~20 \% to above 60 \%, while high-density areas such as Cell-Free MIMO climb from ~40 \% to over 80 \%. 
Regular debate yields uneven 5–10 \% gains and leaves niche gaps while Hierarchical debate achieves near‐uniform high coverage. 
These per‐category insights reinforce our earlier findings: multi‐stage decomposition and targeted sub-task refinement can achieve the breadth, depth, and consistency required for reliable, comprehensive coverage in technical question answering and complex task planning.

In addition, to understand how the number of debate rounds affects our pipeline. We conducted experiments on varying numbers of debate rounds for the strongest model configuration (GPT-4o with GPT-4o-mini). 
Table \ref{tab:debate_rounds_comparison} investigates the effect of increasing debate rounds and reveals a clear pattern of diminishing returns. In the Regular Debate setting, a second refinement pass drops MCR from 49.75 \% to 47.07 \%, and a third round further erodes performance to 45.13 \%. We believe this stems from cascading context drift: each additional pass risks overwriting high-precision terms with marginal or spurious additions, ultimately introducing noise rather than substantive gains. 

\begin{figure}[!t]
  \centering

  \subfigure[Hierarchical Debate]{ \includegraphics[width=0.48\textwidth,height=0.3\textheight,keepaspectratio]{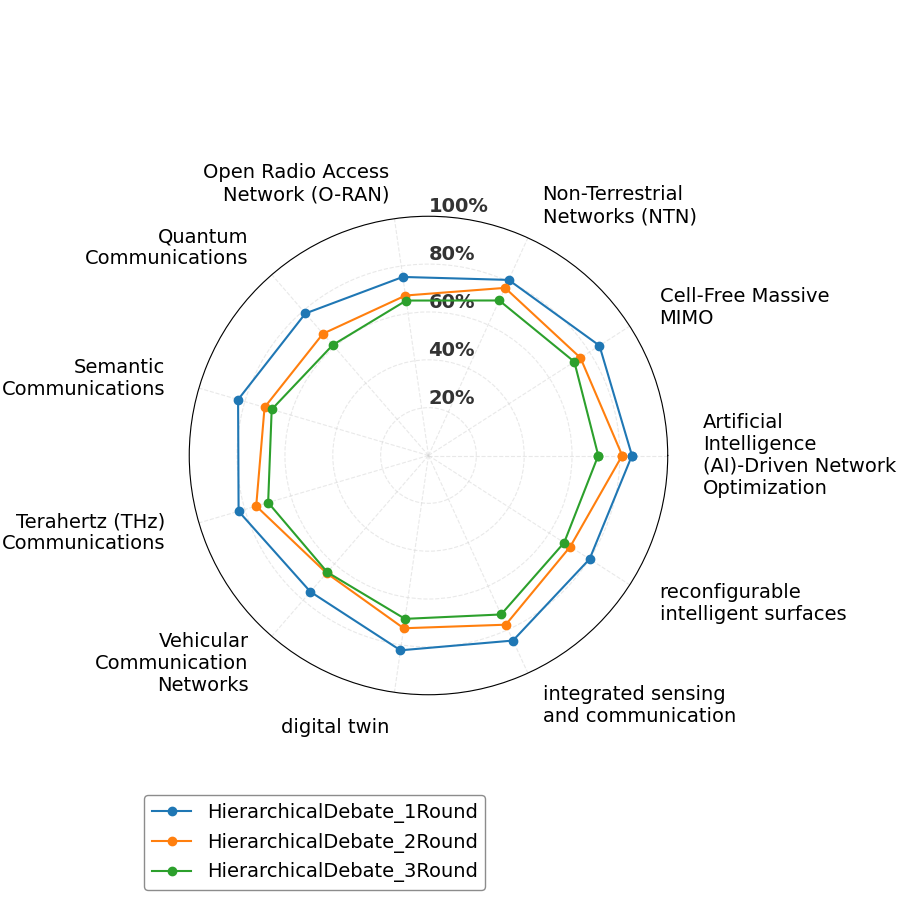}%
    \label{fig:radar-4o-3.5-turbo-round}%
  }
  \subfigure[Regular Debate]{ \includegraphics[width=0.48\textwidth,height=0.3\textheight,keepaspectratio]{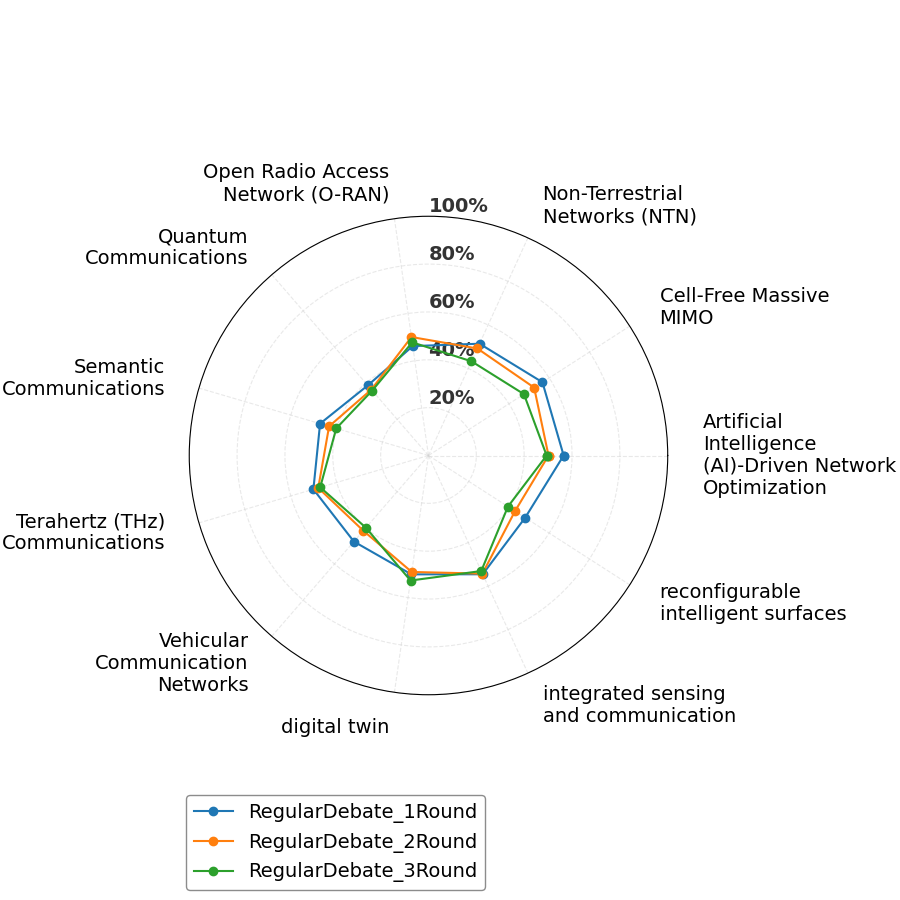}%
    \label{fig:radar-4o-4omini-round}%
  }

  \caption{MCR over 11 technical categories for the GPT-4o + GPT-4o-mini configuration with varying number of debate rounds.}
  \label{fig:radar-combined2}
  \vspace{-15pt}
\end{figure}

\begin{table}[t]
  \centering
  \caption{Impact of Debate Round Count on GPT-4o + GPT-4o-mini Performance for Regular and Hierarchical Debate}
  \label{tab:debate_rounds_comparison}
  \small
  \setlength{\tabcolsep}{6pt}
  \begin{tabular}{llccc}
    \toprule
    \textbf{Debate Type} & \textbf{Rounds} & \textbf{MCR (\%)} & \textbf{KHC} & \textbf{GRR (\%)} \\
    \midrule
    \multirow{3}{*}{Regular}     & 1 Round  & \textbf{49.75} & \textbf{45.87} & \textbf{49.33} \\
                                 & 2 Rounds & 47.07 & 43.56 & 46.85 \\
                                 & 3 Rounds & 45.13 & 41.75 & 44.90 \\
    \midrule
    \multirow{3}{*}{Hierarchical}& 1 Round  & \textbf{81.19} & \textbf{75.41} & \textbf{81.09} \\
                                 & 2 Rounds & 72.83 & 67.54 & 72.63 \\
                                 & 3 Rounds & 68.49 & 63.58 & 68.37 \\
    \bottomrule
  \end{tabular}
\vspace{-15pt}
\end{table}
This pattern mirrors the over-correction effects reported by Chen et al., who demonstrate that excessive multi-round refinement can introduce noise and reduce overall reasoning quality [\cite{chen2024magicore}].
Furthermore, refinement rounds merely produced synonyms of existing keywords, adding no new approaches or algorithms.
Hierarchical Debate follows a similar trajectory—MCR falls from 81.19 \% after one top-level plus sub-task cycle to 72.83 \% and 68.49 \% after two and three rounds. Here, over-decomposition into 20–30 sub-tasks per question fragments the model’s attention, leading to low-value or redundant sub-tasks that dilute overall MCR (See Appendix \ref{sec:app-sampleoutput} for sample output). 
Figure \ref{fig:radar-combined2}’s radar plots reinforce this finding at the category level. Both Regular and Hierarchical Debate profiles contract markedly with each additional round: the one-round trace forms the largest, while the two- and three-round traces shrink and become increasingly jagged. In Regular Debate, well-covered domains like Cell-Free MIMO lose 5–8 \% per extra round, and niche areas such as Semantic Communications collapse from ~40 \% to below 25 \%. Hierarchical Debate suffers even sharper per-category drops—up to 15 \% in Terahertz Communications—reflecting how over-segmentation into 20–30 sub-tasks fragments focus. Together, Table \ref{tab:debate_rounds_comparison} and Figure \ref{fig:radar-combined2} demonstrate that a single, carefully constrained debate round strikes the optimal balance: additional rounds compound noise and undermine both breadth and consistency.

\section{Conclusion}

LLMs are promising technologies to enable AI-native 6G networks, contributing to better understanding of complex network architecture and diverse techniques. 
This work proposed a novel hierarchical debate-based method for complex task planning in the 6G domain. It first decouples these problems into more manageable sub-tasks, and then debates each sub-task to improve the technical details. 
The experiments demonstrate that the proposed technique can outperform conventional debate techniques over various metrics.

\newpage
\bibliography{References.bib}
\bibliographystyle{icml2025}

\newpage
\appendix
\onecolumn

\section{6GPlan Dataset Samples}
\label{datasetsample-app}
The 6GPlan dataset includes the following categories: Reconfigurable Intelligent Surfaces (RIS), Integrated Sensing and Communication (ISAC)
mmWave and Terahertz (THz) Communications, Non-Terrestrial Networks (NTN)
Cell-Free Massive MIMO, Artificial Intelligence (AI)-Driven Network Optimization, Semantic Communications, Open Radio Access Network (O-RAN)
Quantum Communication for 6G Blockchain for Secure Wireless Networks,
6G-Enabled Digital Twin Network. These 11 techniques are widely considered promising techniques towards 6G networks.\\ 
The 6GPlan dataset is available at \url{https://github.com/haozhou1995/6GPlan_Dataset.git}.

For each category, we have designed 10 complex network optimization/management-related tasks. Considering “\textbf{Artificial Intelligence (AI)-Driven Network Optimization}” as an example, the questions include: 

\begin{itemize}
\begin{itshape}
    \item  1.	How can AI-driven systems balance real-time network optimization with long-term infrastructure planning under dynamic traffic conditions?
(Integrates short-term adaptability with strategic resource allocation.)

    \item  2.	How to design AI models that dynamically optimize heterogeneous network resources (e.g., spectrum, power, compute) while ensuring fairness across users?
(Addresses multi-objective trade-offs and fairness constraints.)

    \item  3.	How can reinforcement learning frameworks be structured to handle non-stationary network environments with unpredictable user behavior?
(Focuses on RL robustness against distributional shifts and adversarial conditions.)

    \item  4.	How to implement federated learning for distributed network optimization without compromising latency or data privacy?
(Balances decentralized AI training with QoS and security requirements.)

    \item  5.	How should AI-driven network management systems prioritize conflicting objectives (e.g., energy efficiency vs. ultra-low latency) in 5G/6G slicing?
(Requires Pareto-optimal solutions for multi-dimensional optimization.)

    \item  6.	How to architect AI-based self-organizing networks (SONs) that minimize human intervention while avoiding catastrophic misconfigurations?
(Focuses on fail-safes and interpretability in autonomous decision-making.)

    \item  7.	How can generative AI models simulate and preemptively optimize network topologies for unanticipated traffic patterns?
(Leverages synthetic data and scenario generation for proactive planning.)

    \item  8.	How to integrate digital twin frameworks with AI-driven optimization for cross-domain network lifecycle management?
(Combines simulation, real-time analytics, and closed-loop control.)

    \item  9.	How can transfer learning reduce the cost of deploying AI optimization models across geographically diverse networks?
(Addresses domain adaptation and knowledge reuse in heterogeneous environments.)

    \item  10.	How to quantify and mitigate risks of AI-driven optimization decisions destabilizing legacy network protocols?
(Ensures backward compatibility and graceful degradation during AI adoption.)
\end{itshape}
\end{itemize}

Then, for question 1, “How can AI-driven systems balance real-time network optimization with long-term infrastructure planning under dynamic traffic conditions?”, the dataset is organized as: 

"question50": \{ "question":" How can AI-driven systems balance real-time network optimization with long-term infrastructure planning under dynamic traffic conditions?",\\
             "Answer": "AI-driven network optimization, dynamic traffic prediction, reinforcement learning, digital twins, multi-objective optimization, LSTM networks, spatial-temporal analysis, federated learning, software-defined networking (SDN), network function virtualization (NFV), load balancing, edge computing, generative adversarial networks (GANs), capacity forecasting, 5G network deployment, time-series forecasting, online learning, explainable AI (XAI), OSS/BSS integration, QoS metrics, adversarial training, spectrum allocation, predictive maintenance, graph-based optimization, synthetic data generation, feedback loops, cost-benefit analysis, anomaly detection, regulatory compliance, real-time network optimization, long-term infrastructure planning, dynamic traffic conditions, network efficiency, dynamic resource allocation, machine learning, centralized control, predictive analytics, capacity planning, simulation and modeling, hierarchical AI systems, continuous learning, adaptive learning mechanisms, edge intelligence, network slicing, energy efficiency optimization, latency minimization, security-aware optimization.",
              "Category": "Artificial Intelligence (AI)-Driven Network Optimization" 
\}

Consider another question “How can reinforcement learning frameworks be structured to handle non-stationary network environments with unpredictable user behavior?”, and the dataset sample is: 

"question52": \{ "question":" How can reinforcement learning frameworks be structured to handle non-stationary network environments with unpredictable user behavior?",
             "Answer": "Reinforcement Learning (RL) frameworks, AI-Driven Network Optimization, non-stationary environments, dynamic networks, user behavior variability, channel conditions, network topology, bandwidth allocation, power management, time-varying demands, data rate optimization, adaptive state representation, online feature engineering, real-time channel quality indicators, network load metrics, user mobility patterns, data normalization/scaling, context-aware state augmentation, dynamic reward shaping, multi-objective optimization, adaptive exploration strategies, concept drift detection, meta-learning, transfer learning, domain adversarial training, federated learning, Markov decision processes (MDPs), multi-agent systems, distributed reinforcement learning, hybrid models, Bayesian networks, time-series analysis, anomaly detection, clustering algorithms, contextual bandits, safe exploration, edge computing, network traffic patterns, probabilistic modeling, synthetic perturbations, automated retraining, time-split validation, fallback policies, adaptive learning rates, performance benchmarks.",
                "Category": "Artificial Intelligence (AI)-Driven Network Optimization" 
\}

\begin{figure*}[h]
\centering
\includegraphics[width=0.9\linewidth]{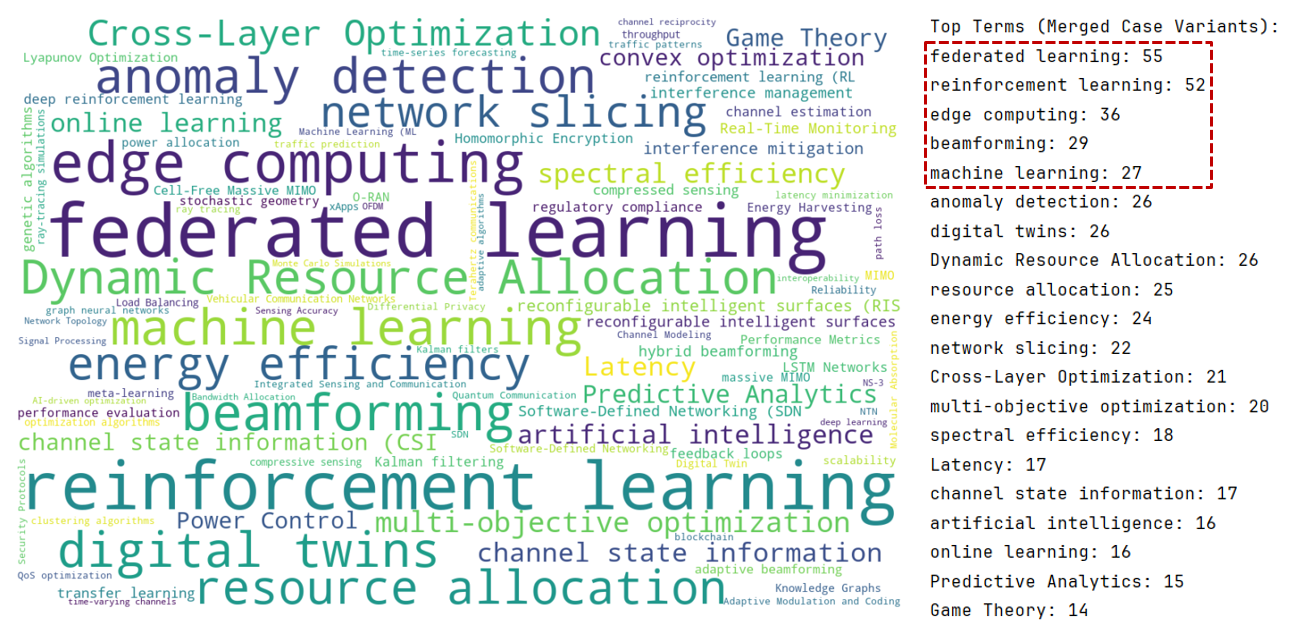}
\caption{Overview of keywords frequencies in the 6GPlan dataset.}
\label{fig-plan}
\end{figure*}

Finally, Fig. \ref{fig-plan} visualized the frequency of different words in the created dataset. It highlights the importance of machine learning techniques for network management, e.g., federated learning, reinforcement learning, edge computing, etc.

\newpage
\section{Prompt Input}
\begin{tcolorbox}[title={Prompt input of LLMs - One Round}]
\label{sec:prompt_1}
\small
\setlength{\parskip}{4pt}
\setlength{\parindent}{0pt}

{\large\bfseries 1. Baseline Pipeline}
\begin{quote}\itshape
“You are an expert in Category: \texttt{\{category\}}.
Given a technical question in this category, Question: \texttt{\{question\}}.
List all relevant technical keywords, methods, algorithms, and designs.
No extra explanation; provide your keyword-rich solution.”
\end{quote}

\medskip
{\large\bfseries 2. Regular Debate}\par
\begin{enumerate}
  \item \textbf{Debater 1 (GPT-4o)}\\
    \begin{quote}\itshape
    “You are Debater 1 (GPT-4o), an expert in Category: \texttt{\{category\}}.
    Given Question: \texttt{\{question\}}, list the relevant technical keywords, methods, algorithms, and designs without extra explanation.
    Provide your keyword-rich solution.”
    \end{quote}
  \item \textbf{Debater 2 (GPT-4o-mini)}\\
    \begin{quote}\itshape
    “You are Debater 2 (GPT-4o-mini).
    Given the same Category: \texttt{\{category\}} and Question: \texttt{\{question\}}, read Debater 1’s solution: \texttt{\{sol\}}.
    Enrich this list by adding any missing technical keywords, methods, algorithms, or designs.
    No extra explanation; provide your keyword-rich solution.”
    \end{quote}
\end{enumerate}

\medskip
{\large\bfseries3. Hierarchical Debate}\par
\medskip
\textbf{3.1 Top-Level Decomposition}\par
\begin{enumerate}
  \item \textbf{Debater 1 (GPT-4o)}\\
    \begin{quote}\itshape
    “You are an expert in \texttt{\{category\}} research.
    Break down the following technical question into a flat, numbered list of high-level steps.
    Question (Category: \texttt{\{category\}}): \texttt{\{question\}}.
    Focus on stages such as prediction, system modeling, optimization, evaluation, etc.
    Do not use nested lists.”
    \end{quote}
  \item \textbf{Debater 2 (GPT-4o-mini)}\\
    \begin{quote}\itshape
    “You are an expert in \texttt{\{category\}} research.
    Review the plan below and insert or refine any missing steps.
    Category: \texttt{\{category\}} \quad Question: \texttt{\{question\}}.
    Initial Decomposition: \texttt{\{sol\}}.
    Keep the numbering; no extra explanation.”
    \end{quote}
\end{enumerate}
\medskip
\textbf{3.2 sub-task-Level Debate}\par
\begin{enumerate}
  \item \textbf{Debater 1 (GPT-4o)}\\
    \begin{quote}\itshape
    “You are Debater 1 (GPT-4o), an expert in \texttt{\{category\}}. Given Question: \texttt{\{question\}}, focus exclusively on sub-task: \texttt{\{st\}}. List the techniques, algorithms, and designs that address this sub-task.\\
    No extra explanation.”
    \end{quote}
  \item \textbf{Debater 2 (GPT-4o-mini)}\\
    \begin{quote}\itshape
    “You are Debater 2 (GPT-4o-mini), focusing on sub-task: \texttt{\{st\}} in Category: \texttt{\{category\}}.
    Given Question: \texttt{\{question\}}, read Debater 1’s answer: \texttt{\{sol\}}.
    Enrich it with any missing techniques, algorithms, or designs.
    No extra explanation.”
    \end{quote}
\end{enumerate}

\end{tcolorbox}

\newpage
\begin{tcolorbox}[title={Prompt input of LLMs - Multiple Rounds}]
\small
\setlength{\parskip}{4pt}
\setlength{\parindent}{0pt}

\medskip
{\large\bfseries Regular Debate (2-3 Rounds)}\par
\begin{enumerate}
  \item \textbf{Round 1: Debater 1 (GPT-4o)}\\
    \begin{quote}\itshape
    “You are Debater 1 (GPT-4o), an expert in Category: \texttt{\{category\}}.  
    Given a technical question, Question: \texttt{\{question\}}.  
    List the relevant technical keywords, methods, algorithms, and designs without extra explanation. Provide your keyword-rich solution.”  
    \end{quote}

  \item \textbf{Round 1: Debater 2 (GPT-4o-mini)}\\
    \begin{quote}\itshape
    “You are Debater 2 (GPT-4o-mini).  
    Given the same Category and Question, read Debater 1’s solution: \texttt{\{sol\}}.  
    Enrich with any missing technical keywords, methods, algorithms, or designs without explanation. Provide your keyword-rich solution.”  
    \end{quote}

  \item \textbf{Round 2: Debater 1 (GPT-4o)} – refine Debater 2’s list\\
    \begin{quote}\itshape
    “You are Debater 1 (GPT-4o).  
    Read Debater 2’s list: \texttt{\{sol\}}.  
    Add any missing technical keywords, methods, algorithms, or designs without explanation. Provide your keyword-rich solution.”  
    \end{quote}

  \item \textbf{Round 2: Debater 2 (GPT-4o-mini)} – refine Debater 1’s update\\
    \begin{quote}\itshape
    “You are Debater 2 (GPT-4o-mini).  
    Read Debater 1’s refined list: \texttt{\{sol\}}.  
    Add any missing technical keywords, methods, algorithms, or designs without explanation. Provide your keyword-rich solution.”  
    \end{quote}

  \item \textbf{Round 3: Debater 1 (GPT-4o)} – further refine\\
    \begin{quote}\itshape
    “You are Debater 1 (GPT-4o).  
    Read Debater 2’s refined list: \texttt{\{sol\}}.  
    Add any missing technical keywords, methods, algorithms, or designs without explanation. Provide your keyword-rich solution.”  
    \end{quote}

  \item \textbf{Round 3: Debater 2 (GPT-4o-mini)} – final refine\\
    \begin{quote}\itshape
    “You are Debater 2 (GPT-4o-mini).  
    Read Debater 1’s final list: \texttt{\{sol\}}.  
    Add any missing technical keywords, methods, algorithms, or designs without explanation. Provide your keyword-rich solution.”  
    \end{quote}
\end{enumerate}

{\large\bfseries Hierarchical Top-level Debate (2-3 Rounds)}\par
\begin{enumerate}
  \item \textbf{Debater 1 (GPT-4o)}\\
    \begin{quote}\itshape
    “You are an expert in \texttt{\{category\}} research.
    Break down the following technical question into a flat, numbered list of high-level steps.
    Question (Category: \texttt{\{category\}}): \texttt{\{question\}}.
    Focus on stages such as prediction, system modeling, optimization, evaluation, etc.
    Do not use nested lists.”
    \end{quote}
  \item \textbf{Debater 2 (GPT-4o-mini)}\\
    \begin{quote}\itshape
    “You are an expert in \texttt{\{category\}} research.
    Review the plan below and insert or refine any missing steps.
    Category: \texttt{\{category\}} \quad Question: \texttt{\{question\}}.
    Initial Decomposition: \texttt{\{sol\}}.
    Keep the numbering; no extra explanation.”
    \end{quote}
  \item \textbf{Round i (alternating GPT-4o / GPT-4o-mini)}\\
\\  
  \itshape i = \# of Debate Rounds * 2.
    \begin{quote}\itshape
    “You are an expert in \texttt{\{category\}} research.
    Review the plan below and insert or refine any missing steps.
    Category: \texttt{\{category\}} \quad Question: \texttt{\{question\}}.
    Previous Decomposition: \texttt{\{sol\}}.
    Keep the numbering; no extra explanation.\\
    {\bfseries Repeat this prompt for 2–3 rounds, feeding each new \texttt{\{sol\}} back into the next iteration, or until no further changes are suggested.}
    ”
    \end{quote}
\end{enumerate}
\end{tcolorbox}

\newpage
\begin{tcolorbox}[title={Prompt input of LLMs - Multiple Rounds}]
\small
\setlength{\parskip}{4pt}
\setlength{\parindent}{0pt}
\medskip

\medskip
{\large\bfseries Hierarchical sub-task-Level Debate (2-3 Rounds)}\par
\begin{enumerate}
  \item \textbf{Round 1: Debater 1 (GPT-4o)} – initial sub-task list\\
    \begin{quote}\itshape
    “You are Debater 1 (GPT-4o), focusing on sub-task: \texttt{\{st\}} in Category: \texttt{\{category\}}.  
    Given Question: \texttt{\{question\}}, list the techniques, algorithms, and designs that address this sub-task. No extra explanation.”  
    \end{quote}

  \item \textbf{Round 1: Debater 2 (GPT-4o-mini)} – first refine\\
    \begin{quote}\itshape
    “You are Debater 2 (GPT-4o-mini), focusing on sub-task: \texttt{\{st\}}.  
    Given Question: \texttt{\{question\}}, read Debater 1’s answer: \texttt{\{sol\}}.  
    Enrich with any missing techniques, algorithms, or designs. No extra explanation.”  
    \end{quote}

  \item \textbf{Round 2: Debater 1 (GPT-4o)} – second refine\\
    \begin{quote}\itshape
    “You are Debater 1 (GPT-4o), focusing on sub-task: \texttt{\{st\}}.  
    Given Question: \texttt{\{question\}}, read Debater 2’s answer: \texttt{\{sol\}}.  
    Enrich with any missing techniques, algorithms, or designs. No extra explanation.”  
    \end{quote}

  \item \textbf{Round 2: Debater 2 (GPT-4o-mini)} – third refine\\
    \begin{quote}\itshape
    “You are Debater 2 (GPT-4o-mini), focusing on sub-task: \texttt{\{st\}}.  
    Given Question: \texttt{\{question\}}, read Debater 1’s refined answer: \texttt{\{sol\}}.  
    Enrich with any missing techniques, algorithms, or designs. No extra explanation.”  
    \end{quote}

  \item \textbf{Round 3: Debater 1 (GPT-4o)} – fourth refine\\
    \begin{quote}\itshape
    “You are Debater 1 (GPT-4o), focusing on sub-task: \texttt{\{st\}}.  
    Given Question: \texttt{\{question\}}, read Debater 2’s refined answer: \texttt{\{sol\}}.  
    Enrich with any missing techniques, algorithms, or designs. No extra explanation.”  
    \end{quote}

  \item \textbf{Round 3: Debater 2 (GPT-4o-mini)} – final refine\\
    \begin{quote}\itshape
    “You are Debater 2 (GPT-4o-mini), focusing on sub-task: \texttt{\{st\}}.  
    Given Question: \texttt{\{question\}}, read Debater 1’s final answer: \texttt{\{sol\}}.  
    Enrich with any missing techniques, algorithms, or designs. No extra explanation.”  
    \end{quote}
\end{enumerate}

\end{tcolorbox}

\newpage
\section{Sample output of Hierarchical Debate}
\label{sec:app-sampleoutput}

Here we provide a sample question output regarding the sub-task decomposition of hierarchical debate. 
The sample question is "\textbf{\textit{How to optimize RIS placement in 3D urban environments to maximize coverage while minimizing blockage effects?}}"
The following are the sub-task decomposition results under different debate rounds. 
The results show that the increased number of rounds may add noise to the sub-task decomposition, e.g., the number of sub-tasks increases from 13 to 23. 
This pattern shows the over-correction effects, e.g., excessive multi-round refinement can introduce noise and reduce overall reasoning quality in debate process.

\textbf{1) Top-level debated sub-tasks for 1 round of debate}
\begin{enumerate}[label=\textbf{Sub-task \arabic*:}, left=5pt]
  \item Conduct a comprehensive analysis of the urban 3D environment to gather data on building layouts, heights, and materials.
  \item Develop a detailed 3D model of the urban area using Geographic Information System (GIS) data and building blueprints.
  \item Identify key locations for potential RIS placement considering existing infrastructure and strategic points for coverage.
  \item Model the electromagnetic wave propagation characteristics in the environment using ray tracing or similar techniques.
  \item Simulate various RIS configurations to understand their impact on signal coverage and blockage in the 3D model.
  \item Formulate an optimization problem to maximize coverage and minimize blockage, defining objective functions and constraints based on the environment and RIS capabilities.
  \item Implement optimization algorithms, such as genetic algorithms or particle swarm optimization, tailored to the problem’s constraints and objectives.
  \item Evaluate candidate solutions from the optimization process using performance metrics like signal strength, coverage area, and interference levels.
  \item Perform sensitivity analysis to assess the robustness of the optimized RIS configurations against variations in environmental parameters.
  \item Validate the optimized RIS placement through field testing or advanced simulation tools to ensure practical feasibility and performance.
  \item Refine the model and optimization criteria based on feedback from real-world testing to improve accuracy and reliability.
  \item Explore alternative RIS technologies and configurations to enhance performance.
  \item Document the entire process, including assumptions, methodologies, and results, to facilitate further research and development.
\end{enumerate}

\textbf{2) Top-level debated sub-tasks for 2 rounds of debate}
\begin{enumerate}[label=\textbf{Sub-task \arabic*:}, left=5pt]
  \item Develop a software tool or platform for continuous monitoring and re-optimization of RIS placements using real-time data.
  \item Incorporate machine learning techniques to enhance the prediction model's accuracy and adaptability to dynamic urban environments.
  \item Ensure compliance with regulatory standards and obtain necessary approvals for RIS deployment in urban areas.
  \item Collaborate with telecommunication providers to integrate RIS placements into existing network infrastructure.
  \item Establish a data collection protocol to gather performance metrics and environmental changes over time.
  \item Implement security measures to protect the RIS infrastructure from potential cyber threats or unauthorized access.
  \item Train personnel on the installation, maintenance, and management of the RIS system to ensure efficient operation.
  \item Develop user-friendly interfaces for stakeholders to visualize and interact with coverage and blockage data.
  \item Explore partnerships with technology companies to leverage advancements in RIS and related technologies.
  \item Set up a feedback loop with end-users to continuously gather insights and improve RIS performance and user experience.
  \item Conduct a comprehensive site survey to identify potential RIS locations based on coverage gaps and obstruction analysis.
  \item Utilize simulation software to model different RIS configurations and their impact on signal propagation in 3D urban environments.
  \item Perform a cost-benefit analysis for various RIS placement scenarios to evaluate financial viability and resource allocation.
  \item Develop algorithms for optimal RIS placement considering parameters such as building heights, density, and user distribution.
  \item Pilot test selected RIS configurations in a controlled urban area to assess performance and gather empirical data.
  \item Analyze the impact of environmental factors (e.g., weather, foliage) on RIS performance and adjust deployment strategies accordingly.
  \item Create a maintenance schedule and protocol to ensure ongoing performance evaluation and infrastructure upkeep.
  \item Document lessons learned and best practices from initial deployments to inform future RIS optimization efforts.
  \item Collaborate with urban planners to integrate RIS deployment into future urban development projects for sustainable growth.
  \item Engage in community outreach to educate local residents about the benefits and functionality of RIS technology.
\end{enumerate}

\textbf{3) Top-level debated sub-tasks for 3 rounds of debate}
\begin{enumerate}[label=\textbf{Sub-task \arabic*:}, left=5pt]
  \item Define the objectives of RIS placement, including coverage maximization and blockage minimization.
  \item Gather and process 3D urban environment data, including building layouts and heights.
  \item Model the radio frequency (RF) propagation characteristics in urban environments.
  \item Identify potential RIS locations based on the 3D environment data.
  \item Develop a mathematical model to represent the coverage and blockage effects.
  \item Formulate an optimization problem to maximize coverage and minimize blockage, including constraints.
  \item Select an appropriate optimization algorithm suitable for the formulated problem.
  \item Implement the optimization algorithm to determine optimal RIS placement.
  \item Simulate the radio wave propagation with the proposed RIS placement using RF simulation software.
  \item Analyze the simulation results to evaluate coverage performance and blockage reduction.
  \item Refine the placement strategy based on evaluation results and iterate if necessary.
  \item Validate the optimized RIS placement through real-world experiments or further simulations with realistic parameters.
  \item Incorporate dynamic factors such as user mobility and varying traffic patterns into the model.
  \item Evaluate the impact of different RIS technologies (e.g., passive vs. active) on performance parameters.
  \item Conduct sensitivity analysis on parameters affecting RF propagation, RIS effectiveness, and environmental variables.
  \item Explore multi-objective optimization techniques to balance coverage, cost-effectiveness, and deployment complexity.
  \item Investigate the integration of machine learning methods to predict optimal RIS placement based on historical data.
  \item Conduct a feasibility study for RIS deployment considering regulatory, infrastructural, and economic constraints.
  \item Document the findings and recommendations for future RIS deployment in urban environments.
  \item Develop a risk assessment plan to identify potential challenges and mitigation strategies during deployment.
  \item Create a comprehensive stakeholder engagement strategy to address community concerns and regulatory compliance.
  \item Set up a monitoring and evaluation framework to assess the long-term performance of the deployed RIS.
  \item Plan for scalability and adaptability of the RIS network to accommodate future technology advancements and urban development.
\end{enumerate}

\end{document}